\documentclass[aps,prl,twocolumn,superscriptaddress,showpacs]{revtex4}

\usepackage{graphicx,color}
\usepackage{dcolumn}
\usepackage{bm}

\begin{document}


\title{Strong spin resonance on BaFe$_2$(As$_{0.65}$P$_{0.35}$)$_2$ 
with $T_c$ = 30 K}



\author{Motoyuki Ishikado}
\email[Corresponding author:]{ishikado.motoyuki@jaea.go.jp}
\affiliation{Quantum Beam Science Directorate, Japan Atomic Energy Agency, Tokai, Ibaraki 319-1195, Japan}
\affiliation{Nanoelectronics Research Institute, National Institute of Advanced Industrial Science and Technology, Tsukuba, Ibaraki 305-8562, Japan}
\affiliation{JST, Transformative Research-Project on Iron Pnictides (TRIP), Chiyoda, Tokyo 102-0075, Japan}

\author{Yuki Nagai}
\affiliation{CCSE, Japan  Atomic Energy Agency, 6-9-3 Higashi-Ueno, Tokyo 110-0015, Japan}
\affiliation{CREST(JST), 4-1-8 Honcho, Kawaguchi, Saitama, 332-0012, Japan}
\affiliation{JST, Transformative Research-Project on Iron Pnictides (TRIP), Chiyoda, Tokyo 102-0075, Japan}

\author{Katsuaki Kodama}
\affiliation{Quantum Beam Science Directorate, Japan Atomic Energy Agency, Tokai, Ibaraki 319-1195, Japan}
\affiliation{JST, Transformative Research-Project on Iron Pnictides (TRIP), Chiyoda, Tokyo 102-0075, Japan}

\author{Ryoichi Kajimoto}
\affiliation{J-PARC Center, Japan Atomic Energy Agency, Tokai, Ibaraki 319-1195, Japan}
\affiliation{JST, Transformative Research-Project on Iron Pnictides (TRIP), Chiyoda, Tokyo 102-0075, Japan}

\author{Mitsutaka Nakamura}
\affiliation{J-PARC Center, Japan Atomic Energy Agency, Tokai, Ibaraki 319-1195, Japan}

\author{Yasuhiro Inamura}
\affiliation{J-PARC Center, Japan Atomic Energy Agency, Tokai, Ibaraki 319-1195, Japan}

\author{Shuichi Wakimoto}
\affiliation{Quantum Beam Science Directorate, Japan Atomic Energy Agency, Tokai, Ibaraki 319-1195, Japan}
\affiliation{JST, Transformative Research-Project on Iron Pnictides (TRIP), Chiyoda, Tokyo 102-0075, Japan}

\author{Hiroki Nakamura}
\affiliation{CCSE, Japan  Atomic Energy Agency, 6-9-3 Higashi-Ueno, Tokyo 110-0015, Japan}
\affiliation{CREST(JST), 4-1-8 Honcho, Kawaguchi, Saitama, 332-0012, Japan}
\affiliation{JST, Transformative Research-Project on Iron Pnictides (TRIP), Chiyoda, Tokyo 102-0075, Japan}

\author{Masahiko Machida}
\affiliation{CCSE, Japan  Atomic Energy Agency, 6-9-3 Higashi-Ueno, Tokyo 110-0015, Japan}
\affiliation{CREST(JST), 4-1-8 Honcho, Kawaguchi, Saitama, 332-0012, Japan}
\affiliation{JST, Transformative Research-Project on Iron Pnictides (TRIP), Chiyoda, Tokyo 102-0075, Japan}

\author{Katsuhiro Suzuki}
\affiliation{Department of Applied Physics and Chemistry, The University of Electro-Communications, Chofu, Tokyo 182-8585, Japan}
\affiliation{JST, Transformative Research-Project on Iron Pnictides (TRIP), Chiyoda, Tokyo 102-0075, Japan}

\author{Hidetomo Usui}
\affiliation{Department of Applied Physics and Chemistry, The University of Electro-Communications, Chofu, Tokyo 182-8585, Japan}

\author{Kazuhiko Kuroki}
\affiliation{Department of Applied Physics and Chemistry, The University of Electro-Communications, Chofu, Tokyo 182-8585, Japan}
\affiliation{JST, Transformative Research-Project on Iron Pnictides (TRIP), Chiyoda, Tokyo 102-0075, Japan}

\author{Akira Iyo}
\affiliation{Nanoelectronics Research Institute, National Institute of Advanced Industrial Science and Technology, Tsukuba, Ibaraki 305-8562, Japan}
\affiliation{JST, Transformative Research-Project on Iron Pnictides (TRIP), Chiyoda, Tokyo 102-0075, Japan}

\author{Hiroshi Eisaki}
\affiliation{Nanoelectronics Research Institute, National Institute of Advanced Industrial Science and Technology, Tsukuba, Ibaraki 305-8562, Japan}
\affiliation{JST, Transformative Research-Project on Iron Pnictides (TRIP), Chiyoda, Tokyo 102-0075, Japan}

\author{Masatoshi Arai}
\affiliation{J-PARC Center, Japan Atomic Energy Agency, Tokai, Ibaraki 319-1195, Japan}
\affiliation{JST, Transformative Research-Project on Iron Pnictides (TRIP), Chiyoda, Tokyo 102-0075, Japan}

\author{Shin-ichi Shamoto}
\affiliation{Quantum Beam Science Directorate, Japan Atomic Energy Agency, Tokai, Ibaraki 319-1195, Japan}
\affiliation{JST, Transformative Research-Project on Iron Pnictides (TRIP), Chiyoda, Tokyo 102-0075, Japan}


\date{\today}

\begin{abstract}

We performed inelastic neutron scattering on powder sample of the P-doped iron-based superconductor BaFe$_2$(As$_{0.65}$P$_{0.35}$)$_{2}$ with $T_c = 30$~K, whose superconducting (SC) order parameter is expected to have line nodes.  We have observed spin resonance at $Q \sim 1.2~$\AA$^{-1}$ and $E=12$~meV in the SC state.  The resonance enhancement, which can be a measure of the area of sign reversal between the hole and electron Fermi surfaces (FSs), is comparable to those of other iron-based superconductors without line nodes.  This fact indicates that the sign reversal between the FSs is still dominant in this system, and the line nodes should create only limited area of sign-reversal on a single FS. Hence the system can hold relatively high-$T_c$. Comparison with theoretical calculation indicates horizontal line nodes may be a candidate to reproduce the observation.

\end{abstract}


\maketitle


In the study of unconventional superconductors, neutron scattering has played significant roles in characterizing the antiferromagnetic (AF) spin fluctuation, as well as in determining the symmetry of the order parameter. In particular, a spin resonance mode, observed as an enhancement of the dynamical structure factor $S(Q,E)$ ($Q$: Momentum, $E$: Energy) below superconducting transition temperature ($T_c$), is taken as a direct evidence for the sign inversion of the order parameter. In the case of cuprates, observation of the spin resonance mode is associated with the $d$-wave character of the order parameter \cite{Rossat, Fong, He}. In the case of iron-based superconductors, the spin resonance mode is observed as an enhancement of $S(Q,E)$ in the superconducting (SC) states relative to normal state at $Q_{AF}$=(1/2 1/2) in tetragonal notation \cite{Christianson, Lumsden, Chi, Li, Christianson2, Mook, Qiu}. This observation is considered as an evidence for the $s_{\pm}$ wave superconductivity \cite{Mazin, Kuroki, Korshunov, Maier2}.

Through the research of iron-based superconductors, the symmetry of the SC order parameter has been systematically examined by various methods, including angle-resolved photoemission spectroscopy, penetration depth and thermal conductivity measurements. In most of the iron-based superconductors with relatively high-$T_c$, such as SmFeAs(O,F) ($T_c$=45 K), (Ba,K)Fe$_2$As$_2$ ($T_c$=38 K), PrFeAsO$_{1-y}$ ($T_c$=35 K) and LiFeAs ($T_c$=16 K), experimental results indicate the fully-gapped symmetry \cite{Malone, Hashimoto2, Hashimoto3, Kim}. On the other hand, line node symmetry is proposed in lower-$T_c$ counterparts, such as LaFePO$_{1-y}$ ($T_c$=7 K) and KFe$_2$As$_2$ ($T_c$=3 K) \cite{Fletcher, Hashimoto4}. The tendency apparently suggests that the nodal order parameter is unfavorable for high-$T_c$. However, this tendency turns out to be non-universal, since recent penetration depth, thermal conductivity and NMR relaxation rate measurements on the BaFe$_2$(As,P)$_2$ superconductor have demonstrated that this material possesses line nodes in the order parameter, in spite of its relatively high-$T_c$ of 30 K \cite{Hashimoto, Nakai}. Here the urgent question one should answer is whether relevant SC order parameter of high-$T_c$ superconductivity in BaFe$_2$(As,P)$_2$ is essentially the same or distinct from other high-$T_c$ colleagues. According to several theoretical researches, it is pointed out that SC gap symmetry of P-doped superconductor may change from fully-gapped $s_{\pm}$ wave to nodal $s_{\pm}$ wave or $d$-wave, resulting in new spin resonance at different wave vector \cite{Kuroki2, Graser, Maier}. Inelastic neutron scattering is a proper probe to detect such magnetic response, namely, it may exhibit different magnetic excitations from that of other iron-based high-$T_c$ superconductors.

In this study, we have performed inelastic neutron scattering to characterize the magnetic excitation and the SC gap symmetry of BaFe$_2$(As$_{0.65}$P$_{0.35}$)$_2$ on a $\sim$ 36 g powder sample with $T_c$=30 K. Here the SC gap symmetry is discussed based on the observed features of the resonance peak. One might suspect that the momentum resolution is significantly limited compared to the measurements using single crystal samples because our experiment has been carried out on a powder sample. However, as we shall see, strong resonant behavior as well as its temperature dependence have allowed us to extract essential information. We also note that the first observation of the spin resonance was reported on a powder sample \cite{Christianson}. In this letter, we show that the order parameter of BaFe$_2$(As$_{0.65}$P$_{0.35}$)$_2$ has predominantly $s_{\pm}$ wave symmetry as in other high-$T_c$ iron-based superconductors, and the region where the sign of the order parameter is reversed exists in the limited parts on a single FS. These observations suggest that relatively high-$T_c$(=30 K) superconductivity despite line node symmetry in BaFe$_2$(As,P)$_2$ is also driven by the AF spin fluctuation at $Q_{AF} \sim$(1/2 1/2).



A polycrystalline sample of BaFe$_2$(As$_{0.65}$P$_{0.35}$)$_2$ was prepared by the following method. First, we prepared precursors of BaAs and Fe$_2$As$_{0.3}$P$_{0.7}$ powders using Ba chips, As grains, P grains and Fe powders as starting materials. BaAs was obtained by reacting Ba chips and As grains in an evacuated quartz tube at 500 ${}^\circ\mathrm{C}$ for 10 hrs and then 600 ${}^\circ\mathrm{C}$ for 10 hrs. Fe$_2$As$_{0.3}$P$_{0.7}$ was obtained by reacting the starting materials in a proper molar ratio in an evacuated quartz tube at 500 ${}^\circ\mathrm{C}$ for 10 hrs and then 900 ${}^\circ\mathrm{C}$ for 10 hrs. Then, synthesized precursors were ground at 1:1 ratio using an agate mortar, and pressed into pellets. They were then sintered at 950 ${}^\circ\mathrm{C}$ for 24 hrs in an evacuated quartz tube. All the processes were done in a globe box purged with helium gas. Tetragonal lattice parameters $a$=3.9240 $\AA$ and $c$=12.807 $\AA$ were determined by powder X-ray diffraction pattern, and $T_c$ was determined to be 30 K by SQUID magnetometer under a magnetic field of 5 Oe. These are consistent with the literature values expected for the present P concentration of $\sim$0.35 (optimally-doped) \cite{Jiang, Kasahara}.

Inelastic neutron scattering measurement was performed on a $\sim$ 36 g powder sample, using the Fermi chopper spectrometer 4SEASONS in J-PARC. In the condition of the present measurement, we employed the multi-$E_i$ method \cite{Nakamura} with incident neutron energies of $E_i$ =151.4, 45.5, 21.6, 12.6, and 8.2 meV, simultaneously. The measuring time is 19 hrs at the beam power of 110 kW. Precise temperature ($T$)-dependent measurements for particular momentum-energy $(Q, E)$ point were taken using Triple Axis Spectrometer (TAS-1) installed at the research reactor JRR-3M. Collimation of B-80'-S-80'-80' (B,S denotes blank, sample) and fixed final neutron energy at $E_f$=30.5 meV were utilized.

In Figs. 1(a) and (b), we show two-dimensional (2D) plots of dynamical structure factor $S(Q,E)$ of BaFe$_2$(As$_{0.65}$P$_{0.35}$)$_2$ measured using incident energy of 45.5 meV at 5 and 32 K, respectively. The most salient feature is a bright spot centering around $Q\sim1.2~\AA^{-1}$ and $E \sim$10-15 meV. When the temperature is decreased below $T_c$ (5 K), the feature becomes more pronounced. Scattering vector $Q \sim 1.2~\AA^{-1}$ roughly corresponds to 1.13 $\AA^{-1}$ at $Q=(1/2~1/2~0)$. 
Note that slight deviation of the $Q$ value is caused by powder averaging effect \cite{Ishikado}. It is therefore likely that the observed peak corresponds to the spin excitation at the AF wave vector $Q_{AF} \sim (1/2~1/2)$. Indeed the observed $Q$ position coincides with that of the spin wave excitation observed in the parent BaFe$_2$As$_2$ powder sample \cite{Ewings}. Furthermore, as elaborated below, this feature is similar to the spin resonance mode observed in other iron-based high-$T_c$ superconductors such as (Ba,K)Fe$_2$As$_2$, Ba(Fe,Co)$_2$As$_2$, Ba(Fe,Ni)$_2$As$_2$ and Fe(Se,Te) \cite{Christianson, Lumsden, Chi, Li, Christianson2, Mook, Qiu}.

\begin{figure}[h]
\begin{center}
\includegraphics[width=8.5cm,clip]{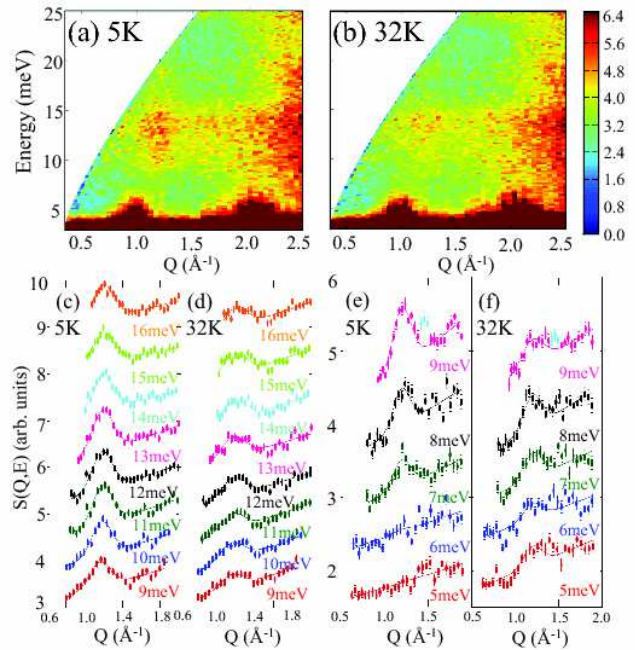}
\caption{(Color online) Dynamical structure factor $S(Q,E)$ of BaFe$_2$(As$_{0.65}$P$_{0.35}$)$_2$ measured at $E_i$= 45.5 meV below $T_c$ (5K:(a)) and above $T_c$ (32K:(b)). Constant energy cuts of $S(Q,E)$ at low temperature (5 K) and just above $T_c$ (32K) for (c), (d) $E_i$=45.5 meV, and (e), (f) $E_i$=21.6 meV. Spurious peaks excluded in the fitting are shown by light blue points.}
\label{Figure1}
\end{center}
\end{figure}

Next, we performed the constant-$E$ cuts of the 2D $S(Q,E)$ plots in order to examine the detailed $E$- and $Q$-dependences. Figures 1(c)-(f) show the constant-$E$ cuts of the $S(Q,E)$ in the energy region of $E$=9-16 meV ((c) and (d)) and $E$=5-9 meV ((e) and (f)), taken at $T$=5 K ((c) and (e)), and 32 K ((d) and (f)). Constant-$E$ cuts were obtained by integrating over the energy resolution (3.4, 1.2 meV for $E_i$ =45.5, 21.6 meV respectively \cite{resolution}). At higher $E$s ($E>$ 7 meV), peak structures centering at $Q_{AF} \sim 1.2~\AA^{-1}$ are recognized, which become more pronounced below $T_c$, thus confirming the results presented in Figs. 1(a) and (b). In contrast, for $E<$ 7 meV, the peak seems to disappear below $T_c$, while it is clearly seen above $T_c$. The suppression of the peak intensity at low $E$s is associated with the opening of the SC gap below $T_c$. In order to carry out quantitative analysis, we have fitted the observed peak by a Gaussian function. The results are plotted as solid lines \cite{fitting}. 

Fig. 2 shows the $E$-dependences of the the dynamical spin susceptibility $\chi"(E)$, which are transformed from $Q$-integrated peak intensity, $S(E)$. At 32 K, $\chi"(E)$ is finite for entire $E$-range and exhibits gradual increase with $E$ up to $E$=14 meV. Note that the absolute value of $\chi"(E)$ is comparable with that of normal state of superconducting sample of LaFeAs(O,F) \cite{Wakimoto}. On the other hand, $\chi"(E)$ at 5 K is characterized by a gap below $E$=6 meV and a peak centered at $E$=12 meV.
The peak corresponds to the bright spot observed in Fig. 1(a) and the enhancement is 1.7 times at 12 meV, relative to the normal state value ($T$=32 K). Similar behavior is also seen in Ba(Fe,Co)$_2$As$_2$ and LaFeAs(O,F) \cite{Lumsden,Wakimoto}.


\begin{figure}[h] 
\rotatebox[origin=lb]{-90}{
\includegraphics[width=4.5cm,clip]{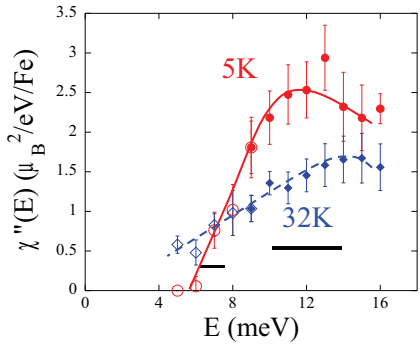}}
\caption{(Color online) Energy dependence of $Q$-integrated dynamical spin susceptibility, $\chi"(E)$, around $Q_{AF} \sim 1.2~\AA^{-1}$ for $E_i$=45.5 meV (filled symbols) and $E_i$=21.6 meV (unfilled symbols) at $T$=5 K (red circles) and $T$=32 K (blue diamonds).Energy resolutions are shown by horizontal bars for double line ($E_i$=21.6 meV) and single thick line ($E_i$=45.5 meV). Solid lines are guides to the eye.}
\label{Figure3}
\end{figure}


In order to observe $T$-evolution more precisely, we have measured the $T$-dependence of the magnetic excitation peak height at $Q=1.2~\AA^{-1}$ and $E$=12 meV by TAS-1. In Fig. 3, the TAS-1 data and the 4SEASONS data are simultaneously plotted. Normalization was done by scaling 4SEASONS data at 10 and 32 K to those of TAS-1. Peak intensity rises up below $T_c$=30 K, confirming that the enhancement of the magnetic excitation 
is closely related with the superconductivity. The most likely origin for the excitation is the spin resonance mode, a signature of 
 the sign inversion in the SC order parameter. We mention that resonance peak energy corresponds to $E_{res}\sim 4.6 k_B T_c$, which well coincides with those of other relatively high-$T_c$ iron-based superconductors \cite{Shamoto}.

\begin{figure}[h]
\rotatebox[origin=lb]{-90}{
\includegraphics[width=4.5cm]{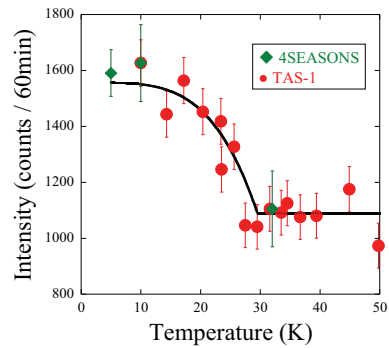}}
\caption{(Color online) Temperature dependence of magnetic excitation peak height observed at $Q=1.2~\AA^{-1}$ and $E$=12 meV. Filled circles represent measured data points for TAS-1 (red filled circles) and 4SEASONS (green diamonds). Black thick line is a guide to the eye.}
\label{Figure4}
\end{figure}

Now let us summarize the experimental results. First, the dominant magnetic excitation appears around $Q_{AF} \sim 1.2~\AA^{-1}$ in the normal state. The absolute value of $\chi"(E)$ is approximately similar to that of LaFeAs(O,F) in the normal state. This observation suggests that magnetic fluctuation arising from $\Gamma$-M nesting dominates in BaFe$_2$(As,P)$_2$. Second, enhancement of $\chi"(E)$ is 1.7 times at 12 meV relative to the normal state value, which is comparable with other systems, such as LaFeAs(O,F) and Ba(Fe,Co)$_2$As$_2$ (almost double \cite{Lumsden, Wakimoto}). Finally, the resonance energy is $E_{res} \sim 4.6k_B T_c$, in agreement with those of the other relatively high-$T_c$ systems. These observation indicates BaFe$_2$(As,P)$_2$ has similar characteristics to other iron-based superconductors with fully-gapped $s_{\pm}$ wave.

The spin resonance in the neutron spectrum is caused via BCS coherence factor:


\begin{equation}
\frac{1}{2} \left(1-\frac{\epsilon_{\bf k} \epsilon_{{\bf k+q}} +\Delta_{\bf k} \Delta_{\bf {k+q}}}{E_{\bf k} E_{{\bf k+q}}} \right)
\approx \frac{1}{2} \left(1-\frac{\Delta_k \Delta_{k+q}}{|\Delta_k| |\Delta_{k+q}|} \right)
\end{equation}
where $E_k =\sqrt{\epsilon_k^2 + \Delta_k^2 }$ is quasi-particle dispersion relation and $\epsilon_k$ is band energy measured relative to Fermi energy. The dynamical spin susceptibility $\chi"(Q_{AF},E)$ in SC state is modified by this coherence factor. 
In $s_{\pm}$-wave case, the signs of the order parameter is inverse ($\Delta_k = -\Delta_{k+Q_{AF}}$) on different FSs, coherence factor becomes almost 1, hence $\chi"(Q_{AF},E)$ is enhanced due to the opening of gap, resulting in the spin resonance. Conversely, when the signs of order parameter is same, coherence factor becomes zero, then $\chi"(Q_{AF},E)$ is suppressed. 
On the other hand, nodal order parameter revealed in BaFe$_2$(As,P)$_2$ by other techniques indicates the existence of region where the sign of order parameter is reversed on a single FS. Such a region should decrease the amount of the enhancement. The observed enhancement ratio of the spectral weight is 1.7 relative to normal state value, which is comparable with about 2 in other systems. 
This suggests that similar amounts of sign reversed process to the case of $s_{\pm}$-wave occurs between $\Gamma$ and M FSs. 
In other words, spin fluctuation mechanism arising from $\Gamma$-M nesting works predominantly like the other relatively high-$T_c$ systems, and it possibly causes high-$T_c$ superconductivity ($T_c$=30 K) despite the existence of nodal symmetry in this BaFe$_2$(As,P)$_2$ system.


Although various symmetries are proposed by theories up to now, the most compatible order pramerter with the experimental results may be that with horizontal line nodes \cite{Graser2,Kurokip}. Actually, we calculate $\chi(Q,E)$ on the basis of the multi-orbital random-phase approximation (RPA) \cite{Kuroki} with use of the effective 10-orbital three-dimensional model at $T = 0$ for various symmetries (i.e. $s_{\pm}$-wave, horizontal node, and $d_{xy}$-wave (node on hole FS)). 
We employ the orbital-interaction coefficients $U_{qt ,{\rm M}}^{rs}$ obtained by the first-principles calculation \cite{Miyake}.
The number of ${\bf k}$-meshes is $512 \times 512 \times 32$. 
We compare the '$s_{\pm}$-wave' model with the gap ($\Delta$) size $\pm 6$ meV 
to the 'horizontal-node' model 
which has the same order parameter as the $s_{\pm}$-wave model except for the horizontal nodes on the most outer hole  Fermi surface. 
%
\begin{figure}
\includegraphics[width = 6cm]{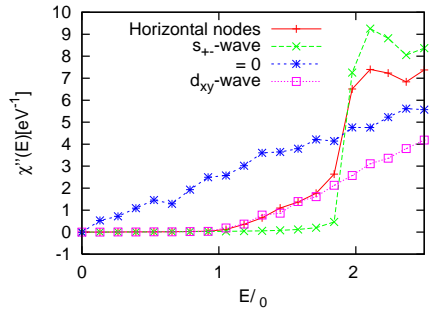}
\caption{\label{fig:spin}
(Color online) Dynamical spin susceptibility $\chi''(E)$ obtained by the multi-orbital RPA with use of the 10-orbital three-dimensional model at $Q \sim (1/2~1/2~1/4)$. 
Energy is scaled by the gap size ($\Delta_0$=6 meV).
}
\end{figure}
As shown in Fig.~\ref{fig:spin}, $\chi''(E)$ in normal states ($\Delta = 0$) at $Q \sim (1/2~1/2~1/4)$ is similar feature to the experimental results \cite{Nagai}. In SC states, the peak intensity ratio relative to normal state around $2 \Delta$ is 2 for $s_{\pm}$ wave and 1.6 for the horizontal-node model. 
 On the other hand, deviation from $s_{\pm}$-wave is large in $d_{xy}$-wave. Therefore, horizontal-node model is well consistent with the experimental results within the experimental accuracy \cite{PocketNode}.



In summary, we have performed inelastic neutron scattering measurement on a powder sample of optimally-doped BaFe$_2$(As$_{0.65}$P$_{0.35}$)$_2$ with $T_c$=30 K. We have observed the characteric magnetic scattering peak around $Q_{AF} \sim 1.2~\AA^{-1}$. The magnetic excitation exhibits pronounced $T$-dependence, namely, gap opening below $E$=6 meV and the enhancement around $E$=12 meV in the SC states. This behavior is regarded as a spin resonance, a signature of unconventional superconductors possessing sign inversion in the order parameter. The enhancement ratio suggests that the order parameter of this system is essentially fully-gapped $s_{\pm}$ wave. Furthermore, combined with the existence of line nodes revealed by other techniques, it is suggested that the region where the sign of the order parameter is reversed exists only in the limited region on a single FS, which causes high-$T_c$ superconductivity ($T_c$=30 K) despite the nodal symmetry in this system.
\\

\begin{acknowledgments}

We thank S. Ishida, K. Kihou, and K. Miyazawa for valuable suggestions. This work was supported by JST, TRIP, and Grant-in-Aid for Specially Promoted Research, Ministry of Education, Culture, Sports, Science and Technology (MEXT), Japan (No. 17001001).

\end{acknowledgments}


\begin{thebibliography}{}
\bibitem{resolution}
We determined integration range as energy resolution by estimating full width at half maximum (FWHM) of incoherent scattering around zero-neutron energy transfer.
\bibitem{fitting}
Here we have employed the peak width $\Delta Q$ (FWHM) at 32 K as 0.29 $\AA^{-1}$ for $E_i$=45.5 meV and 0.25 $\AA^{-1}$ for $E_i$=21.6 meV respectively and fixed the peak position at 1.2~$\AA^{-1}$, since the statistics of the 32K data is not good. As for the results at $E$=9 meV of $E_i$=21.6 meV, three points around $Q=1.4~\AA^{-1}$ are excluded from the fitting since they are spurious peaks due to the instrumental noise.
\end{thebibliography}

{}

\end{document}